\documentclass[aps,pre,preprint,nofootinbib,eqsecnum,tightenlines,
showpacs,amsmath,titlepage]{revtex4}
\usepackage{graphicx}
\global\arraycolsep=2pt
\newcommand{\be}{\begin{equation}}
\newcommand{\ee}{\end{equation}}
\newcommand{\bea}{\begin{eqnarray}}
\newcommand{\eea}{\end{eqnarray}}

\begin{document}

\title{Casimir Energies: Temperature Dependence, Dispersion, and Anomalies}

\author{I. Brevik}
\email{iver.h.brevik@ntnu.no}

\affiliation{Department of Energy and Process Engineering, Norwegian University
of Science and Technology, N-7491 Trondheim, Norway}

\author{K. A. Milton}
\email{milton@nhn.ou.edu}
\affiliation{Oklahoma Center for High Energy Physics and
Department of Physics and Astronomy, The University of Oklahoma,
Norman, Oklahoma 73019, USA}
\date{\today}

\begin{abstract}
Assuming the conventional Casimir setting with two thick parallel
perfectly conducting plates of large extent with a homogeneous and isotropic
medium between them, we discuss the physical meaning of the
electromagnetic field energy $W_{\rm disp}$ when the intervening
medium is weakly dispersive but nondissipative. The presence of
dispersion means that the energy density contains terms of the
form $d[\omega\varepsilon(\omega)] /d\omega$ and
$d[\omega\mu(\omega)] /d\omega$. We find that, as $W_{\rm disp}$
refers thermodynamically to a non-closed physical system, it is
{\it not} to be identified with  the internal thermodynamic energy
$U$ following from the free energy $F$, or the electromagnetic
energy $W$, when the last-mentioned quantities are calculated without
such dispersive derivatives. To arrive at this conclusion, we adopt a
model in which the system is a capacitor, linked to an external
self-inductance $L$ such that stationary oscillations become
possible. Therewith the model system becomes a non-closed one. As
an introductory step, we review the meaning of the
nondispersive energies, $F, U,$ and $W$. As a final topic, we
consider an anomaly connected with local surface divergences
encountered in Casimir energy calculations for higher spacetime
dimensions, $D>4$, and discuss briefly its dispersive
generalization. This kind of application is essentially a
generalization of the treatment of Alnes {\it et al.} [J. Phys. A:
Math. Theor. {\bf 40}, F315 (2007)] to the case of a medium-filled
cavity between two hyperplanes.

\end{abstract}

\maketitle

\section{Introduction}
\label{sec:1} For many years, after its discovery in 1948 \cite{casimir48}
the Casimir effect 
was a theoretical curiosity, although it had evident applications
to van der Waals forces \cite{casimirpolder} and models of hadrons
\cite{bag}.  The Casimir formula for the quantum vacuum force
between conducting plates was generalized to dielectrics by
Lifshitz in 1956 \cite{lifshitz}, and  the 1973 experiment of  Sabisky and
Anderson \cite{sabisky}, testing the Lifshitz prediction to a good
accuracy,  is well known.

But the renaissance in studies of the Casimir effect began in 1997
with the work of Lamoreaux \cite{lamoreaux}.  He measured the
Casimir force between a conducting plate and a spherical lens,
which, through the proximity force approximation
\cite{blocki,deriagin,deriagin2}, agreed with expectations at
something like the 5\% level.  (The accuracy of this measurement
remains under some dispute, because various corrections, such as
the effects of surface roughness, patch potentials, and finite
conductivity, were not adequately taken into account.)  In
subsequent years, a variety of experiments were carried out, some
of much greater accuracy and at considerably shorter distances,
Refs.~\cite{mohideen,roy,harris,ederth,chen,chan,bressi,decca,bezerra,jourdan},
which have incorporated various corrections \cite{bordag}.

One reason why the Casimir effect has attracted so much attention
in recent years is the question of the temperature correction
to the attractive force between real metal surfaces. At large
distances (some micrometers) the relative  thermal
correction achieves several tens of percent, but at these
distances the force itself  becomes weak, and the experimental
technique is not sufficiently sensitive to give clear-cut results.
At small distances, around 100  nanometers, the measurements are
claimed to be of high accuracy, about 1\%, but at such distances
the thermal correction is relatively small.  On the theoretical
side, the process of extracting the temperature dependence was carried
out with the prescription given in Ref.~\cite{schwinger}. Because
of inaccessibility of the effect to precise experiments, the issue
was not reconsidered until the modern era, when Bostr\"om and
Sernelius \cite{bostrom} recognized that this prescription could
not be correct, and that necessarily the transverse electric
reflection coefficient at zero frequency must vanish for metals.
This led to a reduction by a factor of two in the prediction for
the slope of the linear high-temperature behavior (which would
only be visible in experiments carried out at several microns),
but it would predict a new linear temperature term at low
temperatures, resulting in a 15\% correction to the result found
by Lamoreaux.  Lamoreaux believes that his experiment could not be
in error to this extent \cite{lamorpt}. 
Mostepanenko and collaborators have insisted that this behavior is
inconsistent with thermodynamics (the Nernst heat theorem), becaouse
it would predict that the free energy has a term linear in $T$ at low
temperature. Such a behavior is predicted by the so-called modified
ideal metal model and also, as advocated by these authors, when the
Drude model is applied to the case of a metal with perfect crystal
lattice without impurities, in which relaxation of conduction
electrons is only due to scattering on thermal phonons 
\cite{klim01,Bezerra:2007qm,bezerra04}.
Moreover, they assert that the precision Purdue experiments \cite{decca},
performed at $T=300$ K rule out the large thermal corrections
predicted by the use of the Drude model with lattice imperfections
taken into account \cite{hoye03}. The first Purdue experiment was performed at
distances larger than 260 nm. More exact repetitions of that
experiment \cite{decca07}  have been performed, down to 160 nm.
We and others have responded that real metals do not exhibit this
thermodynamic inconsistency, and that most probably the
experiments are not so accurate as claimed
\cite{hoye03,hoye05,hoye06}. The situation is summarized in recent
reviews \cite{Brevik:2006jw,Klimchitskaya:2006rw}.  In particular,
the lack of a thermodynamic inconsistency has been conclusively
demonstrated \cite{Hoye:2007se,Brevik:2007wj}, by showing that the
free energy for a Casimir system made from real metal plates with
impurities has a quadratic temperature dependence at low
temperature. Further evidence for the validity of the notion of
excluding the TE zero mode for metals comes from the recent work
of Buenzli and Martin \cite{buenzli}, corroborating earlier work
by these authors and others \cite{buenzli05,jancovici}, who show
from a microscopic viewpoint that the high-temperature behavior of
the Casimir force is half that of an ideal metal, a rather
conclusive demonstration that the TE zero mode is not
present.\footnote{ It could here be added, as a contrast, that
Intravaia and Henkel have recently claimed that for metals with
perfect crystal lattices the Lifshitz theory leads to violation of
Nernst's theorem \cite{intravaia08}.}

Our purpose with the present paper is not to study  the
temperature corrections in further detail. The brief survey above
indicates that there is a need of reconsidering the underpinnings
of the Casimir theory in some detail. As an attempt to do this, we
will consider the Casimir problem from an unconventional angle,
emphasizing the role of dispersive media. Thus, in
Sec.~\ref{sec:5} we will show how the Casimir energy $W_{\rm disp}$
for a dispersive nondissipative medium, reflecting a {\it
non-closed} physical system, is  not to be identified with the
internal thermodynamic energy $U$, or the electromagnetic energy
$W$, when $U$ and $W$ are calculated as though dispersion were not present. In
this regard a capacitor model of the system proves to be quite
illuminating. Finally, in Sec.~\ref{sec:6} we examine another aspect of
phenomenological electrodynamics, namely its generalization to
higher spacetime dimensions, $D>4$, both because the topic has
some relationship to the dispersive theory discussed in
Sec.~\ref{sec:5}, and also because the higher-dimensional
electrodynamical theory is a topic of general current interest.

\section{Free energy $F$,  internal energy $U$, and electromagnetic energy $W$}
\label{sec:2}

In order to fix the notation, and for reference purposes, we give
in this section a brief survey of how the various energy concepts
occur in Casimir theory. Assume the usual configuration, in which
there are two thick infinitely large plates situated at $z=0$ and
$z=a$, with a homogeneous and isotropic medium in between. We take
this intervening medium to have permittivity $\varepsilon$ and
permeability $\mu$. In this section we take these material
parameters to be constants. For simplicity we assume that the
medium to the left $(z<0)$, as well as that to the right $(z>a)$
are ideal (the so-called IM model), so that the TE and TM
reflection coefficients $r_{\rm TE}$ and $r_{\rm TM}$ satisfy
$r_{\rm TE}^2=r_{\rm TM}^2=1$ for all Matsubara frequencies $m$, including
$m=0$ (the breakdown of this assumption for the $r_{\rm TE}$
coefficient at $m=0$ is the crux of the temperature controversy
for real metals). Let $n=\sqrt{\varepsilon \mu}$ be the refractive
(temperature independent)  index of the intervening medium,
$\beta=1/T,\,\zeta_m=2\pi m/\beta,\,$ and
$\kappa^2=k_\perp^2+n^2\zeta^2$. The free energy $F$ per unit
surface area can now be written
\begin{equation}
F=\frac{1}{\pi \beta}{\sum_{m=0}^{\infty}}{}'
\int_{n\zeta_m}^\infty\kappa d\kappa\ln\left( 1-e^{-2\kappa
a}\right), \label{4}
\end{equation}
for arbitrary $T$. The prime on the summation sign means that the
$m=0$ term is counted with half-weight.

The internal energy per unit area $U$ is now constructed from the
thermodynamical formula
\begin{equation}
U=\frac{\partial (\beta F)}{\partial \beta}. \label{5}
\end{equation}
From Eq.~(\ref{4}) it is apparent that $\beta$ appears only in the
lower limit of the integral in the expression for $\beta F$. Since $\partial
\zeta_m/\partial \beta=-2\pi m/\beta^2$, we get
\begin{equation}
U=\frac{4\pi n^2}{\beta^3}{\sum_{m=0}^\infty}{}'
m^2\ln\left(1-e^{-\alpha m}\right), \label{6}
\end{equation}
where
\begin{equation}
\alpha=\frac{4\pi na}{\beta}=4\pi naT. \label{7}
\end{equation}
The $m=0$ term does not contribute. One way of processing the
expression (\ref{6}) is to expand the logarithm,
\begin{equation}
m^2\ln\left(1-e^{-\alpha m}\right)=-\sum_{k=1}^\infty
\frac{1}{k}m^2e^{-\alpha  km}, \label{8}
\end{equation}
and then sum over $m$, whereby we get
\begin{equation}
U=-\pi n^2T^3\sum_{m=1}^\infty \frac{1}{m}\,\frac{\coth(2\pi
nmaT)}{\sinh^2(2\pi nmaT)}. \label{10}
\end{equation}
When $n=1$, this agrees with Eq.~(18) of Ref.~\cite{feinberg}.
(Cf. also the discussion of energy and free energy in
Ref.~\cite{bezerra02}.) The expansion (\ref{10}) is most
convenient at high temperatures, $aT \gg 1$. By including only the
$m=1$ term, one gets
\begin{equation}
U =-4\pi n^2T^3e^{-4\pi naT},\quad aT \gg 1. \label{11}
\end{equation}
It is apparent that $U \rightarrow 0$ when $T\rightarrow \infty$.
This is as we should expect physically: The Casimir energy
measures the change in energy induced by the boundaries, and these
constraints decrease in importance when the classical thermal
energy becomes high.

To get a convenient expression at low $T$ one may perform a
Poisson resummation, along the same lines as discussed in
Ref.~\cite{schwinger}. Define the quantity $b(m)$,
\begin{equation}
b(m)=m^2\ln\left(1-e^{-\alpha |m|}\right), \label{12}
\end{equation}
along with its Fourier transform $c(q)$,
\begin{equation}
c(q)=\frac{1}{2\pi}\int_{-\infty}^\infty b(x)e^{-iqx}dx.
\label{14}
\end{equation}
Then, according to the Poisson formula,
\begin{eqnarray}
\sum_{m=-\infty}^\infty b(m)&=&2\pi \sum_{m=-\infty}^\infty c(2\pi
m)
=2\int_0^\infty x^2\ln\left(1-e^{-\alpha x}\right)dx\nonumber\\
&&\qquad\mbox{}+4\sum_{m=1}^\infty \int_0^\infty x^2\cos(2\pi mx)\ln
\left(1-e^{-\alpha x}\right) dx. \label{15}
\end{eqnarray}
The various terms can be evaluated analytically. The following
formulas are here useful (Ref.~\cite{Gradshteyn-Ryzhik},
sec.~3.951), assuming $b>0$,
\begin{subequations}
\begin{eqnarray}
\int_0^\infty \frac{x^{2m}\sin bx}{e^x-1}dx&=&(-1)^m
\frac{\partial^{2m}}{\partial b^{2m}}\left[ \frac{\pi}{2}\coth \pi
b-\frac{1}{2b}\right], \label{18}\\
\int_0^\infty \frac{x^{2m+1}\cos bx}{e^x-1}dx&=&(-1)^m
\frac{\partial^{2m+1}}{\partial b^{2m+1}}\left[ \frac{\pi}{2}\coth
\pi b-\frac{1}{2b}\right]. \label{19}
\end{eqnarray}
\end{subequations}
The calculation gives, for arbitrary $T$,
\begin{eqnarray}
U&=&
2\pi n^2T^3\Bigg[ -\frac{\pi}{1440(naT)^3}+\frac{naT}{\pi^3}\sum_{m=1}^\infty
\frac{1}{m^4}\bigg\{-3+\frac{\pi m}{2naT}\coth \frac{\pi m}{2naT}
\nonumber\\
&&\qquad\quad
\mbox{}+\frac{\left(\frac{\pi m}{2naT}\right)^2}{\sinh^2\left(\frac{\pi
m}{2naT}\right)}\left[ 1+\frac{\pi m}{2naT}\coth \frac{\pi
m}{2naT}\right]\bigg\} \Bigg]. \label{20}
\end{eqnarray}
It is of interest to consider the limit of low dimensionless
temperatures,
\begin{equation}
U=-\frac{\pi^2}{720na^3}\left[
1-720\left(\frac{naT}{\pi}\right)^3\zeta(3)+48(naT)^4\right],\quad
aT\ll 1.\label{22}
\end{equation}
Again, this agrees with the low-temperature expression obtained
earlier, for instance in Ref.~\cite{Hoye:2002at}, when $n=1$. It
is to be noted that $U$, as well as the corresponding
low-temperature expression for $F$,
\begin{equation}
F=-\frac{\pi^2}{720na^3}\left[
1+360\left(\frac{naT}{\pi}\right)^3\zeta(3)-(2naT)^4\right],\label{23}
\end{equation}
contain a term that is independent of $a$, which means that this term
does not contribute to the force between the plates.

The third kind of energy that we shall consider is the
electromagnetic energy $W$, still taken per unit surface area. As
above, we take the medium to be nondispersive. We start from the
energy density,
\begin{equation}
w=\frac{1}{2}\varepsilon
(E_z^2+E_\perp^2)+\frac{1}{2}\mu(H_z^2+H_\perp^2), \label{24}
\end{equation}
so that, per unit area, $W=wa$. Quantum mechanically, the product
$E_z^2({\bf r})$ is to be replaced by the expectation value
$\langle E_z({\bf r})E_z({\bf r'})\rangle$ in the limit when ${\bf
r' \rightarrow r}$. Similarly for the other components. We assume
first that $T=0$. According to the fluctuation-dissipation theorem
in Fourier space we have
\begin{subequations}
\begin{eqnarray}
i\langle E_i({\bf r})E_k({\bf r'})\rangle_\omega&=&{\rm Im}\,
\Gamma_{ik}({\bf r,r'};\omega), \label{25}\\
i\langle H_i({\bf r})H_k({\bf
r'})\rangle_\omega&=&\frac{1}{\mu^2\omega^2}\mbox{curl}_{ij}\,
\mbox{curl}'_{kl}\,{\rm
Im}\, \Gamma_{jl}({\bf r,r'};\omega), \label{26}
\end{eqnarray}
\end{subequations}
where $\mbox{curl}_{ik} \equiv \epsilon_{ijk}\partial_j$,
$\epsilon_{ijk}$ being the Levi-Civit\`a symbol. Further, $\bf
\Gamma$ is the Green's function as defined by
Schwinger et al.~\cite{schwinger}, in terms of a polarization source ${\bf P}$,
\begin{equation}
{\bf E}(x)=\int d^4 x' \,{\bf \Gamma}(x,x')\cdot {\bf P}(x'),
\label{27}
\end{equation}
with
\begin{equation}
{\bf \Gamma}(x,x')=\int_{-\infty}^\infty
\frac{d\omega}{2\pi}e^{-i\omega \tau}\,{\bf \Gamma(r,r'};\omega),
\label{28}
\end{equation}
and $\tau=t-t'$. Introducing a transverse Fourier transform,
\begin{equation}
{\bf \Gamma(r,r'};\omega)=\int
\frac{d^2k_\perp}{(2\pi)^2}\,e^{i{\bf k_\perp \cdot (r-r')}}\,
{\bf g}^E(z,z'; {\bf k_\perp},\omega), \label{29}
\end{equation}
we can write
\begin{subequations}
\begin{eqnarray}
g^E_{xx}&=&-\frac{\kappa}{\varepsilon}\frac{1}{d}\cosh \kappa(z-z'),
\label{30}\\
g^E_{yy}&=&\frac{\mu \omega^2}{\kappa}\frac{1}{d}\cosh \kappa (z-z'),
\label{31}\\
g^E_{zz}&=&\frac{k_\perp^2}{\kappa \varepsilon}\frac{1}{d}\cosh \kappa
(z-z'), \label{32}
\end{eqnarray}
\end{subequations}
where
\begin{equation}
d=e^{2\kappa a}-1, \quad \kappa^2=k_\perp^2-n^2\omega^2.
\label{33}
\end{equation}
(Details are given in Ref.~\cite{Ellingsen:2006qh}.) (Note that the
notation is slightly different than that given in Ref.~\cite{mono}.)

Defining the  Fourier components $\langle ..\rangle_{\omega \bf
k}$ of the energy density according to
\begin{equation}
w=\frac{1}{2}\int_{-\infty}^\infty \frac{d\omega}{2\pi}\int
\frac{d^2k_\perp}{(2\pi)^2}\,[\varepsilon \langle
E^2\rangle_{\omega \bf k}+\mu \langle H^2\rangle_{\omega \bf k} ],
\label{34}
\end{equation}
we first obtain for the electric part, letting $z' \rightarrow z$,
\begin{equation}
\frac{1}{2}\varepsilon \langle E^2\rangle_{\omega \bf
k}=\frac{1}{i}\frac{\varepsilon}{2}
(g^E_{xx}+g^E_{yy}+g^E_{zz})=\frac{n^2\omega^2}{i\kappa}\frac{1}{d}.
\label{35}
\end{equation}
Then defining the magnetic counterpart $g_{ik}^H$ to the electric
part $g_{ik}^E$ according to
\begin{equation}
g_{ik}^H=\frac{1}{\omega^2}
\mbox{curl}_{il}
 \, \mbox{curl}'_{km} \, g^E_{lm},
\label{36}
\end{equation}
we obtain by an analogous calculation, in the limit when
$z'\rightarrow z$,
\begin{equation}
\frac{1}{2}\mu \langle H^2\rangle_{\omega \bf
k}=\frac{1}{i}\frac{1}{2\mu}
(g_{xx}^H+g_{yy}^H+g_{zz}^H)=\frac{n^2\omega^2}{i\kappa}\frac{1}{d}.
\label{40}
\end{equation}
The electric and magnetic contributions to the energy are equal,
as we would expect. Adding the expressions (\ref{35}) and
(\ref{40}) and multiplying with $a$ we obtain, at zero
temperature,
\begin{equation}
W=-\frac{n^2a}{\pi^2}\int_0^\infty d\zeta\,\zeta^2\int_0^\infty
\frac{k_\perp dk_\perp}{\kappa d}, \label{41}
\end{equation}
where a  frequency rotation $\omega \rightarrow i\zeta$ has been
performed. This expression can be further processed by introducing
new coordinates $X=k_\perp =\kappa \cos \theta$, $Y=n\zeta =\kappa
\sin \theta$, with $\kappa=\sqrt{k_\perp^2+n^2\zeta^2}$. We get
\begin{equation}
W=-\frac{1}{48\pi^2na^3}\int_0^\infty \frac{z^3
dz}{e^z-1}=-\frac{\pi^2}{720 na^3}, \label{42}
\end{equation}
in accordance with Eqs.~(\ref{22}) and (\ref{23}).

At arbitrary temperature  we get
\begin{equation}
W=-8\pi n^2aT^3\sum_{m=1}^\infty m^2\int_0^\infty \frac{k_\perp
dk_\perp}{\kappa d}, \label{44}
\end{equation}
with $\kappa=\sqrt{k_\perp^2+(2\pi nmT)^2}$. Alternatively, we may
write
\begin{equation}
W=-4\pi n^2T^3\sum_{m=1}^\infty m^2\int_{\alpha m}^\infty
\frac{dz}{e^z -1}, \label{45}
\end{equation}
where $\alpha=4\pi naT$ as before.

At high temperature, $aT \gg 1$, it is easy to check that $W$
agrees with $U$ calculated previously. We approximate the integral
in Eq.~(\ref{45}) by $\int_{\alpha m}^\infty e^{-z} dz=e^{-\alpha
m}$, and so  get
\begin{equation}
W=-4\pi n^2T^3\sum_{m=1}^\infty m^2e^{-\alpha m} \rightarrow -4\pi
n^2T^3 e^{-4\pi naT} \label{46}
\end{equation}
when $m=1$, in agreement with Eq.~(\ref{11}).

We shall not delve further into a detailed study of the equality
between $W$ and $U$ in the case of arbitrary $T$. The equality
should be clear on physical grounds, since we are dealing with a
closed thermodynamical system.

After having given this survey, we have the necessary reference
background for studying the dispersive regime.

\section{On the dispersive case, neglecting dissipation}
\label{sec:5}

As mentioned, our main focus will be on the dispersive case.
Assume first that the medium in the region $0<z<a$ is both
electrically and magnetically frequency dispersive,
$\varepsilon=\varepsilon(\omega), \mu=\mu(\omega)$. The walls are
taken to be perfectly conducting, as before. The total energy
density $w_{\rm disp}$ is known to be
\cite{Landau-Lifshitz,embook}
\begin{equation}
w_{\rm disp}=\frac{1}{2}\int_{-\infty}^\infty \frac{d\omega}{2\pi}\int
\frac{d^2k_\perp}{(2\pi)^2}\left[ \frac{d(\varepsilon
\omega)}{d\omega}\langle E^2\rangle_{\omega \bf k}+\frac{d(\mu
\omega)}{d\omega}\langle H^2\rangle_{\omega \bf k} \right].
\label{47}
\end{equation}
We can write this as a sum of two parts $w_I$ and $w_{II}$, where
$w_I$ is the same expression as in Eq.~(\ref{34}) with
$\varepsilon \rightarrow \varepsilon(\omega),\, \mu \rightarrow
\mu(\omega)$, and where
\begin{equation}
w_{II}=\frac{1}{2}\int_{-\infty}^\infty \frac{d\omega}{2\pi}\omega
\int \frac{d^2k_\perp}{(2\pi)^2}\left[
\frac{d\varepsilon}{d\omega}\langle E^2\rangle_{\omega \bf
k}+\frac{d\mu}{d\omega}\langle H^2\rangle_{\omega \bf k} \right].
\label{48}
\end{equation}
Correspondingly, for the surface densities,
$W_{\rm disp}=W_I+W_{II}$.

The first property to be noted in connection with  Eq.~(\ref{47})
is that it is derived under the assumption of {\it negligible
dissipation}. Some dissipation is always present---this being a
consequence of Kramers-Kronig's relations---but it is a legitimate
approximation  to neglect it except in the neighborhood of
eigenfrequencies in the cavity. This assumption means that the
relaxation frequency in the dispersion relation can be set equal
to zero, and we may adopt the usual dispersion relation for a
dielectric, for simplicity taking it hereafter to be nonmagnetic,
\begin{equation}
\varepsilon(\omega)=1+\frac{\bar\varepsilon-1}{1-\omega^2/\omega_0^2},
\quad \mu=1. \label{49}
\end{equation}
In the case of a general dissipative medium, neither the energy
nor the stress tensor are derivable in terms of
permittivity/permeability alone, and therefore cannot be given in
a general form using macroscopic methods. (This point is discussed
in detail by Ginzburg \cite{ginzburg}.)

Second, it is clear that the expression (\ref{47}) is not
intimately related to the Casimir effect as such. It is more
natural to consider the problem as belonging to classical
electrodynamics, namely  a system of two conducting plates between
which there are stationary electromagnetic oscillations. The
expression (\ref{47}) is actually obtained from the energy balance
equation
\begin{equation}
\bf \nabla \cdot (E\times H) +E\cdot \dot{D}+H\cdot \dot{B}=0.
\label{50}
\end{equation}
(See, for example, Eq.~(7.5) in Ref.~\cite{embook}.)
In order to accumulate electromagnetic energy, one has to consider
oscillations that are not purely monochromatic, but distributed
within a band of frequencies around each eigenfrequency. In
this way external agencies, outside of the plates,  are called
for. It is natural here to regard the system to be a capacitor,
linked to an external appropriately adjusted self-inductance $L$
such that stationary oscillations become possible (external
resistances are forbidden since we omit dissipation). That means,
the plates with the intervening medium is  thermodynamically a
{\it non-closed\/} system. From this we can draw the important
conclusion that the full dispersive energy $W_{\rm disp}$ is not to be
identified with the thermodynamical energy $W=U$ calculated
earlier. The laws of thermodynamics are applicable to {\it closed\/}
systems only.

The mentioned model of a classical electromagnetic non-dissipative
circuit is studied in Ref.~\cite{Landau-Lifshitz}. It is
instructive to consider the salient features of the argument also
here:

Let the charges $Q$ be supplied and withdrawn from the plates with
frequency $\omega$. The self-inductance of the circuit is $L$, as
mentioned, and the  electromotive force we call $\cal E$. The
potential $\phi$ across the plates is determined from the equation
\begin{equation}
\phi={\cal{E}}-L \dot{ J}, \label{51}
\end{equation}
where $J=\dot{Q}$. The frequency of the circuit is
\begin{equation}
\omega=1/\sqrt{LC(\omega)}, \label{52}
\end{equation}
where the capacitance $C(\omega)$ of the capacitor is determined
by $\phi=Q/C(\omega)$. By considering almost monochromatic
quantities [the same kind of argument that led to Eq.~(\ref{47})],
we get, when taking the average over a period,
\begin{equation}
\overline{{{\cal E}J}}=\frac{d}{dt}\left\{
\frac{1}{2}L\overline{J^2}+\frac{1}{2}\frac{d(\omega
C)}{d\omega}\,\overline{\phi^2} \right\}. \label{53}
\end{equation}
The expression  between brackets is the circuit energy. From
$J=-i\omega Q$ and Eq.~(\ref{52}) we get
$\frac{1}{2}L\overline{J^2}=\frac{1}{2}C\,\overline {\phi^2}$ and
so the circuit energy may be written
\begin{equation}\overline{W}_{\rm
circ}=\frac{1}{2\omega}\frac{d(\omega^2C)}{d\omega}\,\overline{\phi^2}.
\label{54}
\end{equation}
This expression, because of the derivative with respect to
$\omega$, is seen to be related to  Eq.~(\ref{47}).

Now consider a small adiabatic displacement of the capacitor
plates. As $\overline{W}_{\rm circ}/\omega$ is an adiabatic invariant,
\begin{equation}
\delta \overline{W}_{\rm circ} =\overline{W}_{\rm circ}\delta
\omega/\omega. \label{55}
\end{equation}
By means of Eq.~(\ref{52}),
\begin{equation}
\frac{ \delta \omega}{\omega} =-\frac{1}{2}\frac{\delta C}{C}.
\label{56}
\end{equation}
The change in $C$ consists of two parts,
\begin{equation}
\delta C=(\delta C)_{\rm st}+\frac{dC}{d\omega}\delta \omega,
\label{57}
\end{equation}
where the first term is the static part and the second term
depends on the frequency change. From Eqs.~(\ref{56}) and
(\ref{57}),
\begin{equation}
\delta C_{\rm st}=-\frac{1}{\omega^2}\frac{d(\omega^2
C)}{d\omega}\,\delta \omega. \label{58}
\end{equation}
When Eq.~(\ref{54}) is substituted in Eq.~(\ref{55}) and
(\ref{58}) is used, $dC/d\omega$ disappears, and we get
\begin{equation}
\delta \overline{W}_{\rm circ}=-\frac{1}{2}\overline{\phi^2}(\delta
C)_{\rm st}=-\frac{1}{2}\frac{\overline{ Q^2}}{C^2}(\delta C)_{\rm st}.
\label{59}
\end{equation}
This is the same expression as one obtains by taking the variation of the
average of the energy $Q^2/2C$ of a thermally insulated capacitor. It
means that when dispersion is present, the electromagnetic stress
tensor contains no derivatives with respect to the frequency. The
argument is general, and is not critically dependent on our choice
of a capacitor model.

When applied to our case, we can thus conclude as follows:
\begin{enumerate}
\item The dispersive energy $W_{\rm disp}$ whose density is given
in Eq.~(\ref{47}) refers thermodynamically to a non-closed system,
and is therefore not to be identified with the internal energy $U$
calculated in Sec.~\ref{sec:2} starting from  the free energy $F$,
or the electromagnetic energy $W$, in the nondispersive case. As
was demonstrated, when $\varepsilon$ and $\mu$ are constants,
$W=U$.  We are still to use the same expressions for $W$ and $U$
when the permittivity and permeability depend on frequency.

\item As for the electromagnetic stress tensor, the derivatives with
respect to $\omega$ are not to be included. That is, the
electromagnetic force can be calculated from Eq.~(\ref{34}) with
$\varepsilon \rightarrow \varepsilon(\omega),\, \mu \rightarrow
\mu(\omega)$.
\end{enumerate}
\vspace{0.5cm} It may finally be noted that by inserting the
simple form (\ref{49}) for $\varepsilon(\omega)$ for a dielectric,
we obtain for the dispersive correction $W_{II}=aw_{II}$ a
divergent expression,
\begin{equation}
W_{II}=\frac{2a(\bar\varepsilon-1)}{\omega_0^2}\int_0^\infty
\frac{d\omega}{2\pi}\frac{\omega^2}{(1-\omega^2/\omega_0^2)^2}\int
\frac{d^2k_\perp}{(2\pi)^2}\langle E^2\rangle_{\omega \bf k};
\label{60}
\end{equation}
cf. Eq.~(\ref{48}).

Another way to to see that the dispersive medium should be treated
without the frequency derivative of the permittivity is to
recognize that the Casimir energy may be derived by a variation
expression \be \frac{\delta
E}A=\frac{i}2\int\frac{d\omega}{2\pi}\frac{d^2k_\perp} {(2\pi)^2}
dz\,\delta\varepsilon(z)g^E_{kk}(z,z,\mathbf{k_\perp},\omega), \ee
which is Eq.~(2.26) of Ref.~\cite{schwinger}.  This starting point
is equivalent to the variational argument recounted in this
section.

\section{Discussion on an anomaly in the Casimir energy for higher  dimensions}
\label{sec:6}

The electromagnetic theory in a continuous medium has in general a
rich structure. Most notable is the fact that (within the commonly
accepted Minkowski theory) the spatial photon momentum is equal to
$ {\bf{k}}=  n\omega \hat{\bf{k}}$, implying that the photon
four-momentum becomes spacelike, $k^\mu k_\mu=(n^2-1)\omega^2 >0$
(we make use of the metric $g_{\mu\nu}=\mbox{diag}(-1,1,1,1)$, and
assume $n$ constant to begin with). Accordingly, there are
inertial systems in which the photon energy becomes negative. A
striking example of this sort is provided by the \v Cerenkov effect:
In the inertial system where the emitting particle is initially at
rest, the recoil kinetic energy of the particle is necessarily
positive. Thus, in order to preserve energy conservation, the
energy of the emitted photon has to be negative. Another example
of a related sort is provided by the so-called anomalous Doppler
effect, occurring when  a quantum particle detector moves
superluminally in the medium. Thus Ginzburg and Frolov
\cite{ginzburg86} studied such kinds of particle detectors  and
showed how the excitation of a detector  uniformly accelerated
in a vacuum with the associated emission of radiation is actually
similar to the radiation occurring in the region of the anomalous
Doppler effect when the detector is moving  superluminally with
constant velocity in the medium. See also the discussion in
Ginzburg's book \cite{ginzburg}. Situations of these kinds were
discussed also by Brevik and Kolbenstvedt, in the case of constant
velocity \cite{brevik88} and for constant acceleration
\cite{brevik89}.

We shall round off our paper not by considering  the
above-mentioned effects  any further, but instead another effect
that has also a bearing on medium electrodynamics, namely the
anomaly that turns up in the case of higher spacetime dimensions,
$D>4$. The anomaly reflects the breaking of conformal symmetry. We
do this because the topic has some relationship to that considered
in Sec.~\ref{sec:5}, and also because it has attracted
 interest recently in the case of a vacuum field.  A
generalization to the medium case thus appears natural. Higher
dimensions, in the context of Casimir theory, were considered long
ago by Ambj{\o}rn and Wolfram \cite{ambjorn83}, but  anomalies of
the type considered below were not studied until recently  by
Alnes {\it et al.} \cite{alnes07}; cf.\ also
Refs.~\cite{alnes07a,alnes06}.

Let us assume, then, that there are two parallel hyperplanes with
separation $a$,  the region $0<z<a$ being filled with an isotropic
medium of refractive index $n=\sqrt{\varepsilon \mu}$. The walls
are assumed perfectly conducting, as before. The anomaly we wish
to consider is present also in the case of zero temperature, so
we shall assume $T=0$ in the following.

The appropriate electromagnetic energy-momentum tensor is the
Minkowski expression, called $S_{\mu\nu}^M$,
\begin{equation}
S_{\mu\nu}^M=F_{\mu\alpha}H_{\nu}{}^{\alpha}-\frac{1}{4}g_{\mu\nu}
F_{\alpha\beta}H^{\alpha\beta};
\label{60a}
\end{equation}
cf., for instance, Refs.~\cite{moller72,brevik79,brevik70}. Here
$F_{\mu\nu}=\partial_\mu A_\nu-\partial_\nu A_\mu$ with
$\mu,\nu=0, 1, 2,...D-1=d$ is the field tensor, whereas $F^{0k}=E_k$ with
$k=1,2,...d$ are the components of the $d$-dimensional electric
field vector $\bf E$. The magnetic induction ($\bf B$ in the
three-dimensional case) does not constitute a vector in the
higher-dimensional case, but is given by the $d(d-1)/2$ components
of the antisymmetric spatial tensor $F_{ik}$. Analogously, the
second $D$-dimensional tensor $H_{\mu\nu}$ occurring in
Eq.~(\ref{60a}) is given by the vector components $H^{0k}=D_k$,
$\bf D$ being the $d$-dimensional induction vector, and by the
$d(d-1)/2$ components of the spatial magnetic field tensor
$H_{ik}$ ($\bf H$ in the three-dimensional case). In analogy with
three-dimensional theory, we assume constitutive relations in the
form $H^{0k}=\varepsilon F^{0k}$ and $F_{ik}=\mu H_{ik}$ also when
$D>4$.

Turning now to physical quantities, it is convenient to start with
the surface pressure $P$ on the hyperplane $z=0$. We  observe that
 the usual  expression for $P$ (cf.
Eq.~(\ref{4})) can easily be  generalized to the case of $d=D-1$
spatial dimensions. Taking into account that there are $(D-2)$
physical degrees of freedom in the field in the cavity, we have
\begin{equation}
F=(D-2)\int_0^\infty \frac{d\zeta}{2\pi}\int \frac{d^{d-1}
k_\perp}{(2\pi)^{d-1}}\ln(1-e^{-2\kappa a}), \label{62}
\end{equation}
where
\begin{equation}
\kappa^2=k_\perp^2+n^2\zeta^2, \quad k_\perp^2 \equiv
k_x^2+k_y^2+...+k_{D-2}^2. \label{63}
\end{equation}
The volume element in momentum space is
$d^{d-1}k_\perp=\Omega_{d-2}\,k_\perp^{d-2}dk_\perp$, where the
solid angle is determined by 
$\Omega_{d-1}=2\pi^{d/2}[\Gamma(d/2)]^{-1}. $ The pressure
$P=-\partial F/\partial a$ can now be written
\begin{equation}
P=-\frac{2(D-2)}{(2\pi)^d}\,\Omega_{d-2}\int_0^\infty d\zeta \int
\frac{\kappa k_\perp^{d-2}dk_\perp}{e^{2\kappa a}-1}. \label{65}
\end{equation}
The double integral over $\zeta$ and $k_\perp$ can be further
processed by introducing polar coordinates, \cite{schwinger}. Again, we
introduce $X=\kappa \cos \theta=k_\perp,\, Y=\kappa \sin \theta=n\zeta$,
satisfying $X^2+Y^2=\kappa^2$. The area element in the $XY$ plane
is $\kappa d\kappa d\theta=ndk_\perp d\zeta$. The  integral
therewith becomes
\begin{equation}
\int_0^\infty d\zeta \int \frac{\kappa
k_\perp^{d-2}dk_\perp}{e^{2\kappa a}-1}
=\frac{1}{n}\int_0^{\pi/2}\cos^{d-2}\theta d\theta\int_0^\infty
\frac{\kappa^d d\kappa}{e^{2\kappa a}-1}. \label{66}
\end{equation}
We now make use of known integral formulas and insert the
expression for $\Omega_{d-2}$, to get
\begin{equation}
P=-\frac{(D-2)(D-1)}{n}\,\frac{\Gamma(D/2)\zeta(D)}{(4\pi)^{D/2}a^D}.
\label{70}
\end{equation}
It ought to be emphasized that this expression was  obtained
without any regularization procedure. The presence of the medium
is seen here to turn up through the  factor $n$ in the
denominator. If $n=1$, including the
 case of a vacuum as well as the case of a ``relativistic'' medium
satisfying $\varepsilon=1/\mu$, the expression reduces to that
derived earlier \cite{ambjorn83}. This result parallels that
obtained in the $T=0$ parts of the energy, cf. Eqs.~(\ref{22}) and
(\ref{23}).  It is nearly identical to the result found in
Ref.~\cite{mono} for the scalar case in $D$ dimensions [Eq.~(2.35)
there], differing only in the  factor $(D-2)/n$.

The electromagnetic field energy density $w$ in the cavity is a
more delicate quantity.  The natural way to calculate $w$ is via
the energy-momentum tensor. This procedure - carried out by Alnes
{\it et al.} in the case of a vacuum cavity \cite{alnes07,alnes06}
- led in the case of metallic boundary conditions to the result
\begin{equation}
w=-\frac{(D-2)\Gamma(D/2)}{(4\pi)^{D/2}a^D}\left[\zeta(D)+\left(\frac{D}{2}-
2\right)f_D\left(\frac{z}{a}\right)\right]\equiv w_1+w_2,\label{71}
\end{equation}
where
\begin{equation}
f_D\left(\frac{z}{a}\right)=\zeta_H\left(D,\frac{z}{a}\right)+
\zeta_H\left(D,1-\frac{z}{a}\right),
\label{72}
\end{equation}
$\zeta_H$ being the Hurwitz zeta function. Note that the first term yields
the pressure (\ref{70}),
\be
-\frac\partial{\partial a}aw_1=P,
\ee
so that the second term in the energy density, $w_2$,which
diverges like $z^{-D}$ close to the surface when $D>4$,
does not contribute to the force between the plates.
It can explicitly seen that written in physical variables this
term is independent of the separation between the plates and
hence does not contribute to the force. This
anomaly can actually be seen to manifest itself in another way if
we go back to  the expression (\ref{60a}) for the energy-momentum
tensor: its trace $S_{\mu}^{M\mu}$ is  {\it nonvanishing} when
$D>4$. Physically, as emphasized in Ref.~\cite{alnes07}, the
divergent self energy of a single surface is related to the lack
of conformal invariance of the electromagnetic Lagrangian for
$D>4$.  All of this is exactly as seen in Ref.~\cite{mono}, Chap.~11,
for the scalar field.

It turns out that the anomaly can be regularized away by
subtracting off the self energy for both plates. Then, the second
term in Eq.~(\ref{71}) is absent, and only the first, finite,
terms in $w$  remains.

As mentioned, these calculations of $w$ were made for the case of
a vacuum cavity. The result was found via a combination of
dimensional and zeta function regularizations \cite{tywonik04}.
The result could be recalculated for a medium cavity, but such a
detailed calculation is hardly justified in view of the simple
occurrence of $n$ in the expression (\ref{70}). In fact,
since in physical units, $w=\hbar c/a^D$ times a function of $D$,
it is clear (for example, Ref.~\cite{embook}, Eq.~(36.12)) all we
have to do to insert a uniform medium between the plates is replace
$c$ by $c/n$; this
shows that the same factor $n$ will appear in the denominator
of the expression for $w$ as it did in $W$ or $P$.
Thus, after regularization, we obtain the relationship
\begin{equation}
P=(D-1)w_1, \label{73}
\end{equation}
which is the same connection as for a vacuum.

Finally, we consider an alternative method for obtaining the
energy $W$ that avoids the field theoretical approach above, and
instead starts by considering the individual photon momenta
directly. The photon momentum in the medium is
$\sqrt{k_\perp^2+\pi^2m^2/a^2}$, and the photon energy is obtained
by dividing this expression by $n$, assuming that $n=$ constant.
Thus we have, still at $T=0$,
\begin{equation}
W=\frac{1}{n}\sum_{m=1}^\infty \int
\frac{d^{d-1}\,k_\perp}{(2\pi)^{d-1}}\,\sqrt{k_\perp^2+\pi^2m^2/a^2}.
\label{74}
\end{equation}
In order to extract a finite expression we have to regularize in
some way, for instance by using an exponential cutoff. The
important point in our context is however that the integral in
Eq.~(\ref{74}), and the sum, are just the same as in a vacuum
field. Thus the influence of the medium turns up only in the
prefactor $1/n$, in accordance with what was found above.

Can this theory be generalized to the dispersive case? Such a
performance is not  quite straightforward, in view of the
complicated form (\ref{47}) for the dispersive energy density.
Some insight can however be obtained from the following argument.
As noted in Sec.~\ref{sec:5}, the electromagnetic stress tensor
does not contain derivatives with respect to the frequency. Thus,
the momentum flux density has the same form as in a nondispersive
medium. We may assume therefore that the photon wave vector is
equal to $n(\omega)\omega$:
\begin{equation}
\sqrt{k_\perp^2+\pi^2 m^2/a^2}= n(\omega)\omega. \label{75}
\end{equation}
When the wave vector is given, this equation can be solved
(numerically) for $n(\omega)$ and $\omega$. Inverting, we find $n$
as a function of $\sqrt{k_\perp^2+\pi^2m^2/a^2}$. Accordingly, we
can write the energy as
\begin{equation}
W=\sum_{m=1}^\infty \int
\frac{d^{d-1}\,k_\perp}{(2\pi)^{d-1}}\,\frac{\sqrt{k_\perp^2+\pi^2m^2/a^2}}{n(\sqrt{k_\perp^2+\pi^2m^2/a^2})}.
\label{76}
\end{equation}
For very high wave numbers, $n\rightarrow 1$, and the integral
reduces for these frequencies to its vacuum counterpart. One
should bear in mind that the expression (\ref{76}) holds only in
an approximate sense, as we have ignored the accumulation of
 energy during the slow building up of the electromagnetic field.

\section{Conclusions}

Some care ought to be taken when dealing with dispersive and
dissipative media. In the general case of arbitrary dispersion
(which implies necessarily dissipation also), the electromagnetic
energy cannot be rationally defined as a thermodynamic quantity at
all. If dispersion is weak, making it possible to ignore the
accompanying dissipation, it is meaningful to  define the electric
energy such that it contains terms of the type $d[\omega
\varepsilon(\omega)]/d\omega$, and similarly for the magnetic
field. In such a case, as we have seen, one should distinguish the
electromagnetic energy $W_{\rm disp}$ from the thermodynamic energies
calculated for nondispersive media and used without the
derivatives on $\varepsilon$ and $\mu$ in the dispersive case
where $n=n(\omega)$. The main reason for the
difference is that in the dispersive case we are dealing with a
non-closed physical system.

The electrodynamic theory of media, especially when dispersion is
included, has a rich structure. As an example of this, we showed
in Sec.~\ref{sec:6} the anomaly turning up when $D>4$, which is
especially interesting when dispersion is present. This shows
the interplay between local surface energy divergences and the
breaking of conformal symmetry. The clarity brought to bear by the
above  analysis will now allow us to understand more fully the
dispersive case, and to some extent also the questions connected
with temperature problems. Moreover, we hope to have contributed
to the understanding of surface energies and their significance.

Finally, we make the following comment. Our electromagnetic
formalism in this paper has been the conventional one, whereby the
basis for calculating stresses on matter is the Abraham-Minkowski
stress tensor (cf., for instance, Ref.~\cite{brevik79}). Now, in a
recent paper Raabe and Welsch \cite{raabe05} have developed a
somewhat unconventional theory for electromagnetic fields in a
medium based upon the Lorentz force, from which they derive a
stress tensor different from the Abraham-Minkowski form. The
Casimir effect was chosen by these authors as the physical
phenomenon to which they applied their proposed  theory. We
merely  mention this novel formulation here;
it would lead us too far from our main purpose to make a detailed
scrutiny of this rather complicated  formulation. A comment on
some  consequences of the altered stress tensor in practical
applications is under preparation \cite{brevik08a}.

\begin{acknowledgments}
K.A.M.'s research is supported in part by a grant from the US
National Science Foundation (PHY-0554926) and by a grant from the
US Department of Energy (DE-FG02-04ER41305). I. B. thanks Finn
Ravndal and his group at the University of Oslo for valuable
information.  We are grateful to Simen Ellingsen for helpful
conversations.
\end{acknowledgments}

\end{document}